# Evidence of Formation of Superdense Nonmagnetic Cobalt

Nasrin Banu[1], Surendra Singh[2], B. Satpati[3], A. Roy[4], S. Basu[2], P. Chakraborty[3], Hema C. P. Movva[4], V. Lauter[5] and B. N. Dev[1]*

[1]Department of Materials Science, Indian Association for the Cultivation of Science, 2A & 2B Raja S. C. Mullick Road, Jadavpur, Kolkata 700032, India.

[2]Solid State Physics Division, Bhabha Atomic Research Centre, Mumbai 400085, India.

[3]Surface Physics and Material Science Division, Saha Institute of Nuclear Physics, 1/AF, Bidhannagar, Kolkata 700064, India.

[4]Microelectronic Research Center, The University of Texas at Austin,10100 Burnet Road, Bldg 160, MER 1.606J, Austin, Texas 78758, USA.

[5]Quantum Condensed Matter Division, Neutron Science Directorate, Oak Ridge National Laboratory, One Bethel Valley Road, Oak Ridge, TN 37831-6475, USA.

*e-mail: msbnd@iacs.res.in

Abstract

Magnetism of 3$d$ transition metals at high density has always received wide interest due to existence of magnetism at the Earth's core. For ferromagnetic cobalt, although there is a theoretical prediction that its magnetic moment would vanish when the density increases to about 1.4 times the normal value, so far there is no experimental evidence. We have discovered the existence of ultrathin superdense nonmagnetic cobalt layers in a polycrystalline cobalt thin film. The densities of these layers are about 1.3-1.4 times the normal density of Co. This has been revealed by X-ray reflectometry experiments, which provide a depth profile of the electron scattering length density, and corroborated by polarized neutron reflectometry (PNR) experiments. The superdense Co layer has an fcc structure, unlike hcp structure for bulk Co, as revealed by transmission electron microscopy. The magnetic depth profile, obtained by PNR, shows that the superdense Co layers at the film-substrate interface and near the top of the film are nonmagnetic. The major part of the Co film has the usual density and magnetic moment.

**Introduction**

Interest in magnetism in 3$d$ transition metals, iron, nickel and cobalt, has continued beyond the traditional bulk materials and thin films. Recent investigations have concentrated on two-dimensional monolayers, one-dimensional monatomic metal chains [1] and even on single atoms on surfaces [2]. Even for the bulk material, due to the existence of magnetism at Earth's core, the studies of structure and magnetism of 3$d$ transition metals at reduced volume (under high pressure) have received a great deal of attention [3]. These materials are known to be ferromagnetic in the bulk form. Can these materials be nonmagnetic under some conditions? Almost a century ago nonmagnetic films of iron, nickel and cobalt were reported [4]. These films were produced by sputtering in hydrogen, helium or argon gas environment.

These films showed distension. The distension or swelling was as large as 20% due to incorporation of sputtering gas atoms into the transition metals. Among these studies, there have been elaborate investigations on Co. Although the authors could not pinpoint the reason for the loss of magnetism, they conjectured that it was possibly associated with the changed electron distribution in the metallic atom either due to the influence of chemical combination or simply due to the increased distance of the atoms apart. Heating these films at 350$^o$C for a few minutes restored them to their ferromagnetic state, presumably due to removal of absorbed gas from the sputtering process [4].

In the light of some recent theoretical predictions and experimental results, it may be argued that indeed there is a possibility of formation of nonmagnetic 3$d$ transition metals. Theoretical studies have shown the dependence of magnetic moment of Co on its atomic volume. With decreasing atomic volume (or increasing material density), the magnetic moment of Co decreases, and the magnetic moment of an fcc Co phase nearly vanishes when the atomic volume is reduced by a factor of ~1.4 [5, 6]. In a high pressure experiment on Co an hcp to fcc phase transition has been observed around this atomic volume corresponding to a density of ~ 1.4 times the normal Co density [3]. The authors conjectured that this phase was likely to be nonmagnetic. However, no magnetic investigation on this phase of Co under high pressure has so far been reported. Polycrystalline thin films, on the other hand, can produce a compressive stress in the film [7], thereby increasing the material density by mimicking high pressure conditions. Thus, polycrystalline thin films of Co offers a possibility to observe the high density nonmagnetic Co phase.

Here we report on the observation of a nonmagnetic fcc phase of cobalt, the density of which is about 1.3 - 1.4 times the density of normal hcp cobalt. We observe such a dense nonmagnetic phase in an electron-beam deposited polycrystalline Co thin film on a Si(111) substrate. The evidence of a superdense cobalt phase is obtained from X-ray reflectivity (XRR) experiment. The result is corroborated by Rutherford backscattering spectrometry (RBS), secondary ion mass spectrometry (SIMS) and transmission electron microscopy (TEM) experiments. The magnetic and the chemical depth profile analysis [8-10] from polarized neutron reflectometry (PNR) experiments corroborates the high density and also shows that this superdense cobalt layer is nonmagnetic.

**Superdense cobalt**

The density depth profile of the Co film was investigated with X-ray reflectivity measurements, which probe the electron scattering length density (ESLD) with sub-nanometer depth resolution. The XRR data, theoretical fit and the extracted ESLD depth profile are shown in Fig. 1. For the fit to the data, performed with a uniform Co layer, the model significantly deviates from the experimental data over the whole momentum transfer (Q) range (Figure 1a); the corresponding depth profile of ESLD, which includes the contribution of a thin surface cobalt oxide layer, is depicted in the inset. Fig. 1(b) represents the best fit, which corresponds to the ESLD depth profile shown in the inset. The ESLDs show some distinct features, in particular we reveal that near the top of the Co layer and at the Co/Si interface larger values than that of normal Co are needed to describe the XRR data. The histogram (dashed line) in the inset of Fig. 1(b) shows the individual layers with the corresponding ESLD and layer thickness. The effective ESLD includes the effect of the interface roughness. The ESLD at the mid-region of the Co film (~6.2×10$^{-5}$ Å$^{-2}$) is consistent with normal Co (~6.3×10$^{-5}$ Å$^{-2}$). The ESLD close to the air/film interface of the Co film (~

$8.3\times10^{-5}$ Å$^{-2}$, histogram; effectively ~ $7.7\times10^{-5}$ Å$^{-2}$ due to large roughnesses at the interfaces below and above), is much higher (1.34 times) than that of normal bulk Co. At the top of the Co film (surface), a cobalt oxide layer is present as expected. The ESLD of the two oxides, CoO and $Co_3O_4$, are $4.76\times10^{-5}$ Å$^{-2}$ and $4.56\times10^{-5}$ Å$^{-2}$ respectevly, i.e., less than that of Co. This rules out the identity of the high ESLD layer as CoO or $Co_3O_4$. The thin topmost layer (~ 1 nm), where a smaller ESLD is observed, is likely to be CoO (and/or $Co_3O_4$). So, what is the identity of the layer with an ESLD of ~ $8\times10^{-5}$ Å$^{-2}$? The Co film investigated here is a polycrystalline film(seen in TEM discussed latter). In a polycrystalline film, there is a possibility of formation of grains under compressive strain [7] leading to a high density layer. So there is a distinct possibility that the high ESLD layer is actually a high density Co layer with an average density of about 1.3 times the normal Co density. A similar high density layer at the film/substrate interface (Figure 1b inset), with somewhat less density compared to the top high density layer, may also be high density Co. However, in order to be confident that this is indeed a high density Co layer, we need to rule out the presence of any high density element as an inadvertent contaminant in the sample. This has been done by RBS experiments, the results of which are discussed latter. As obtained from the XRR analysis, the total thickness of the film is 27±2 nm.

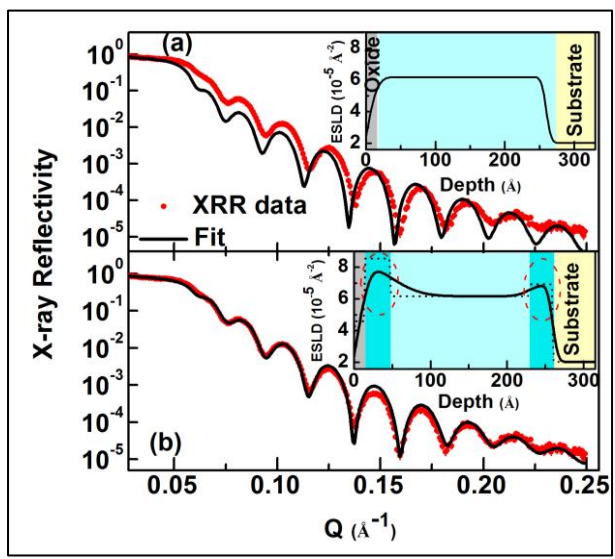

**Figure 1| Observation of higher density layers within the Co film.** X-ray reflectivity from a Co thin film and the electron scattering length density depth profile. (a) XRR data and the theoretical XRR for a constant Co density (or constant ESLD) depth profile as in the inset. Strong deviation of the theoretical curve from the data shows that the uniform density model is unsatisfactory. (b) XRR data and the best fit, the corresponding ESLD depth profile is shown in the inset. High density Co layers are marked deeper green and also by ovals and normal Co in lighter green in the inset.

In order to rule out the possibility of contamination by any high atomic number (Z) high density material, we carried out RBS experiment. The RBS result is shown in Fig. 2(a). Signals from Si and Co are seen in the spectrum. No other element of higher atomic mass (or higher Z) is present in the sample. If they were present they would appear as peak(s) at higher energies than that of the position of Co. It should be mentioned here that under the experimental condition the RBS technique is sensitive to an equivalent of a fraction of one atomic layer for high Z elements (say Au). A simulated RBS spectrum for a small amount of Au (0.5 nm) on the Co/Si sample is shown in the inset of Fig. 2(a) to give an indication of the sensitivity of RBS to high-Z elements. So, the absence of any high-Z element in the RBS spectrum confirms that the high ESLD layer observed in XRR is a high density Co layer.

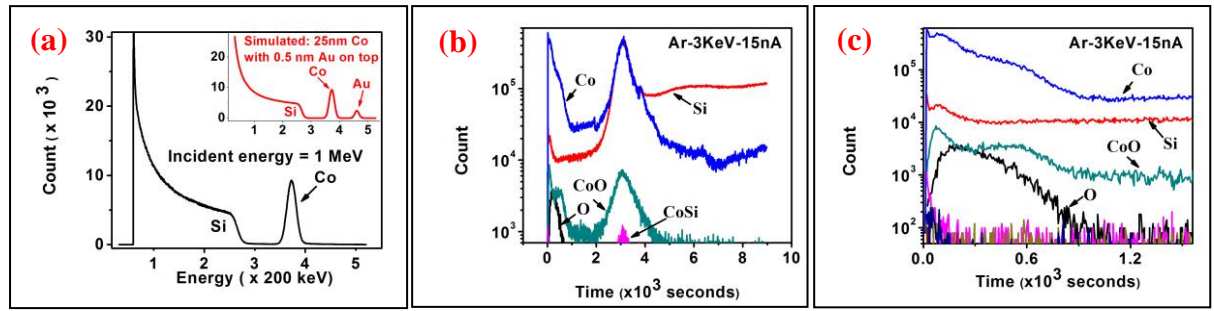

**Figure 2| Absence of higher density element than Co and indication of high density Co**. (a) A RBS spectrum from the sample. The backscattering signal in the energy region ~120-550 keV is from the substrate Si. The peak around 750 keV is from Co. Presence of any high-Z element(s) would have generated signals at energies above 800 keV. For example, the presence of a small amount of Au would have given a spectrum similar to the one shown in the inset. (b), (c) SIMS depth profile of yield of different emitted species (Co, Si, CoO, O; yields of CoSi, $Co_2Si$, $Co_2O_3$, $Co_3O_4$ are very low). Sputtering time is proportional to the depth in the sample. (c) Depth profile closer to the surface; for time larger than ~1000 sec, yield from normal Co is seen. For time less than 1000 sec, higher yield of Co is indicative of high density Co.

We carried out SIMS experiments and detected different sputtered atomic and molecular ions from the sample. Again SIMS data did not show any evidence for any high Z elements. SIMS data are shown in Fig. 2(b-c). In SIMS depth profile, secondary ion-yield of an individual species from the surface and interfaces are seen to get enhanced due to the presence of any impurity element in the matrix (so-called matrix effect). Data from any depth can be safely interpreted provided the matrix effect is properly compensated. In Fig. 2(b-c) the time axis is proportional to depth. In Fig. 2(b) we notice the flat region of the Co yield which comes from the mid-region of the Co film. Ignoring the initial peak at the top surface, arising due to a 'matrix effect' in conventional SIMS, we notice that there is an enhanced Co yield in the outer region of the Co film, which is a probable indication of a high density Co. This feature is more clear from Fig. 2(c), which shows the yield in an expanded time scale up to about $1.5\times10^3$ sec. SIMS data show that there is a very thin layer of CoO at the top of the film and other oxides of Co are practically absent.

All experimental evidences point to the formation of a superdense Co layer. This is further corroborated by PNR results discussed latter. We need to understand the mechanism of its formation and its crystalline phase. A material is expected to have a lower volume or higher density in a high pressure environment. How does a high density phase form in a thin film and retain the higher density at ambient pressure? A high density phase can indeed form in a polycrystalline film. Our Co film is a polycrystalline grainy film. During the growth of a grainy film via physical vapour deposition, deposited atoms could diffuse into grain boundaries producing a compressive stress [7]. This can be a possible reason for the formation of the superdense Co layer. This phenomenon is illustrated in the inset of Fig. 3(a), where *L* is the lateral size of a grain. The flux of deposited atoms is *J* and the growth rate is *dh/dt*. Deposited atoms can diffuse into grain boundaries, thereby producing a compressive stress in the grains. The grains under compressive stress would have a higher density.

Obviously all the grains would not be under the same compressive stress, and consequently there would be a lateral distribution of density. XRR determines the laterally averaged density. Plan view TEM image of our Co film [Fig. 3(a-b)] shows the polycrystalline grainy nature of the film. It also indicates a lateral density distribution in the form of absorption contrast. From Fig. 3(c) we also notice that the superdense Co grains (black grain) have an fcc structure. In Fig. 3(d), from the scanning transmission electron microscopy high-angle annular dark field (STEM-HAADF) images, we notice lateral variation of Co density and the presence of some very high density Co grains (bright). The high density is also reflected in electron energy loss spectroscopy (EELS) spectra from the Co $L_{2,3}$ edges from different grains [Fig.3(e)]. The absorption peak is smaller for the higher density Co grains. This feature has been observed consistently on many bright grains compared to their surroundings. Higher density tends to reduce absorption cross section. From X-ray absorption experiments under high pressure, for several cases it was observed that the intensity of the absorption peak decreases with increasing pressure (i.e., at higher densities) [11, 12]. The drift corrected STEM-HAADF image and the corresponding EDX elemental maps from the orange boxed region in Fig. 3(d) using Co–K, O–K and Si–K energies are shown in Fig. 3(f). EDX map of Co, showing intense fluorescence from some grains compared with their surroundings, also indicates the presence of very high density Co grains.

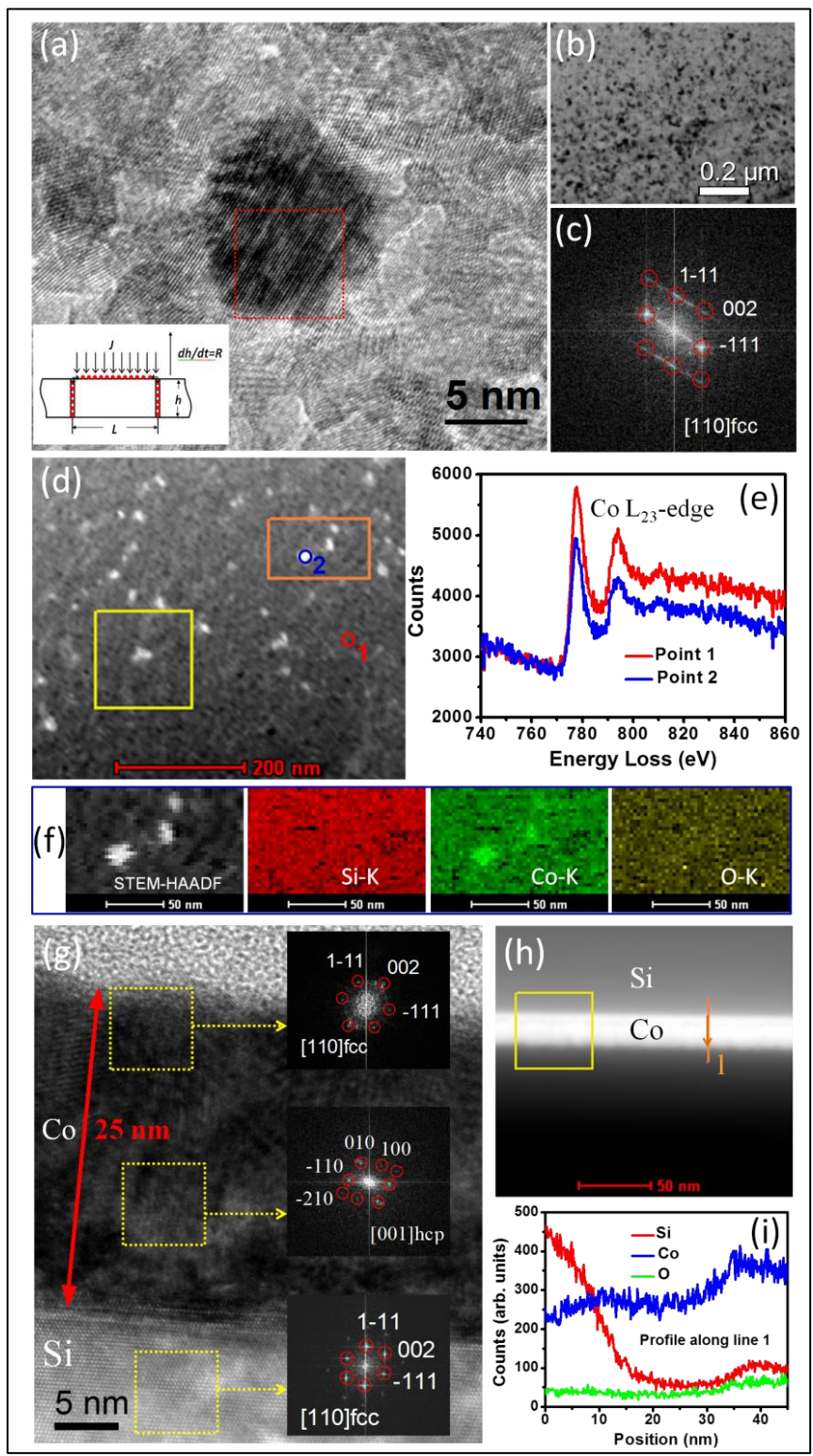

**Figure 3| Observation of high density fcc Co grains and indication of high density Co in depth profile.** (a) Plan view high resolution TEM image. Inset shows the mechanism for the formation of compressed grains in polycrystalline growth by atomic deposition. (Adapted from Ref. 7). (b) A low magnification TEM image. (c) FFT pattern from the marked part of the darker grain shows fcc structure. (d) Plan view STEM-HAADF image. (e) EELS from point 1 and 2 shown in (d). (f) STEM-HAADF image and the corresponding elemental maps from the orange boxed region in (d). (g) XTEM image of about 25 nm Co layer on Si substrate. FFT pattern from different regions (yellow boxes) including the substrate Si are shown in the inset. (h) Cross sectional STEM-HAADF image. (i) Drift corrected X-ray fluorescence yields of Si, Co and O along line 1 in (h).

As mentioned earlier, in a high pressure experiment on Co an hcp to fcc phase transition has been observed around the atomic volume corresponding to a density of ~ 1.4 times the normal Co density [3]. According to a theoretical work [5], the high density nonmagnetic Co is in the fcc phase. Fig. 3 confirms this fcc phase of Co. We have investigated this aspect also by cross-sectional TEM investigation. Fig. 3(g) shows cross-sectional TEM images along with fast Fourier transforms (FFTs) from different regions of the sample. The substrate Si shows the fcc structure, as expected. The mid-depth region of the Co film shows hcp structure and the superdense top layer of the Co film shows fcc structure. It [Fig. 3(i)] also shows the depth profiles of X-ray fluorescence yields of Si, Co and O. The Co fluorescence depth profile indicates the presence of high density Co near the surface of the Co film as well as some density enhancement near the Co/Si interface, consistent with the XRR results.

**Nonmagnetic Cobalt**

Theoretical work has predicted that Co in the fcc phase, at a density of about 1.4 times the normal density, can have a nearly zero magnetic moment [5]. The variation of magnetic moment as a function of volume (or density) is reproduced in Fig. 4(a).

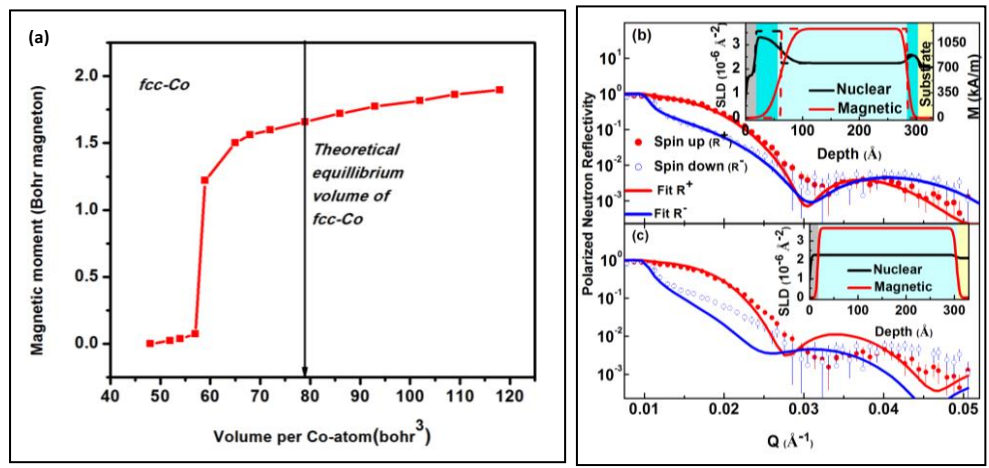

**Figure 4| Dependence of magnetic moment on atomic volume (or material density) for fcc Co**. (a) Atomic volume of normal Co is ~79 bohr$^3$. Around an atomic volume of 57 bohr$^3$ (equivalent to a density of ~ 1.4 times that of normal Co) magnetic moment drastically reduces to approximately zero (adapted from Ref. 4). (b) PNR data and the theoretical fit. The corresponding nuclear and magnetic scattering length density from the fit are shown in the inset. The SLDs are obtained by including the effect of surface and interface roughness. Broad profiles at the superdense-Co/normal-Co interface at a depth of ~ 6 nm is due to large interface roughness (density variation over a few nm). The histograms show the layer structures and layer SLDs. The high density Co layer (nuclear SLD or density ~ 1.4 times the normal Co density) near the top of the Co film has a vanishing magnetic moment. The high density Co layer at the Co/Si interface also has reduced magnetic moment. (c) Uniform Co density, i.e. uniform nuclear and magnetic SLDs as shown in the inset, leads to theoretical reflectivities (R+ and R-) which strongly deviate from the experimental data.

As we have observed such a superdense fcc phase of Co near the top of the Co layer and at the Co/Si interface, we have investigated its magnetic behaviour. Standard macroscopic magnetic measurements would not be able to distinguish the magnetic contribution from this superdense ultrathin Co layers from the magnetic contribution of the entire Co film. So we carried out magnetic depth profiling by PNR experiments [8, 9, 10], on the same sample as the one used in XRR experiment, at the DHRUVA facility at BARC, Mumbai [13]. PNR data and the model fit to the data are shown in Fig. 4(b); the corresponding nuclear and magnetic depth profiles are shown in the inset. It is seen that the Co/Si interface is at a depth of about 30±2 nm. In the mid-region of the depth profiles (inset) the magnetic as well as the nuclear scattering length density (SLD) profile is consistent with normal Co ($\sim 2.26 \times 10^{-6} \text{Å}^{-2}$). Below the Co/Si interface, both the SLDs are consistent with Si. However, from the surface up to a depth of ~ 6 nm, the magnetic SLD is about zero. Out of this 6 nm, the nuclear SLD of the top ~ 1 nm is about $1.7 \times 10^{-6} \text{Å}^{-2}$ while the nuclear SLD ($\sim 3.25 \times 10^{-6} \text{Å}^{-2}$) of the remaining ~ 5 nm is about 1.4 times the nuclear SLD of normal Co. This is consistent with a high density Co observed in XRR. CoO and $Co_3O_4$ have higher nuclear SLDs ($\sim 4.29 \times 10^{-6} \text{Å}^{-2}$ and $\sim 4.69 \times 10^{-6} \text{Å}^{-2}$ respectively) and the measured nuclear SLD ($\sim 3.25 \times 10^{-6} \text{Å}^{-2}$) is much smaller than these values. In addition, ESLD from XRR for this ~ 5 nm thick layer is inconsistent with such cobalt oxides; it is rather consistent with a high density Co layer. Both XRR and PNR show that this ~ 5 nm thick layer is high density Co. PNR additionally shows that the high density Co layer has nearly zero magnetic SLD or magnetic moment density. Similarly the high density Co layer at the Co/Si interface, which has higher NSLD compared to normal Co and consistent with higher ESLD in XRR, also shows nearly zero magnetic moment. Right vertical axis of the the inset in Fig. 4(b) shows the corresponding magnetization in kA/m. The normal Co layer shows a marginally lower value of magnetization (~1250 kA/m) as compared to its bulk value (~ 1400 kA/m). Fig. 4(c) shows the theoretical PNR (R+ and R-) curves for uniform nuclear and magnetic SLDs (inset). The strong deviation of the theoretical curves from the experimental data indicates that the uniform density depth profile of Co is unacceptable.

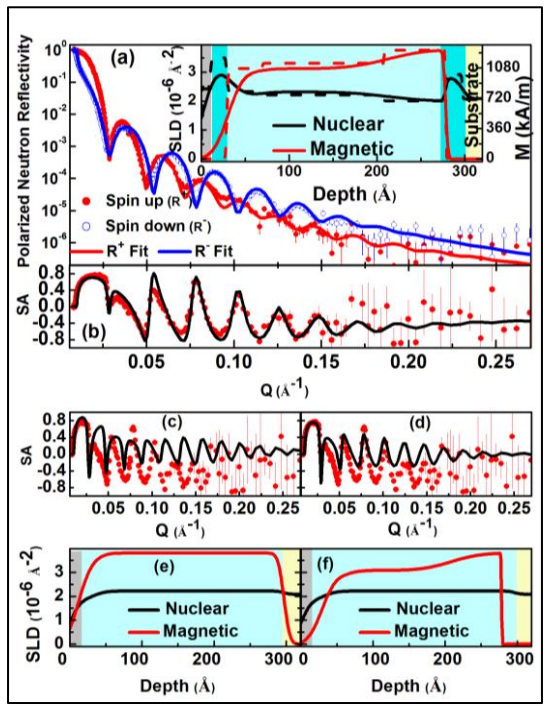

**Figure 5| Additional confirmation of high density Co layers by PNR measurement over a large range of momentum transfer**. (a) PNR data and theoretical fit. Inset: the extracted nuclear and magnetic SLD depth profile including the effect of surface and interface roughness. The corresponding SLD histograms are also shown in the inset. High density Co layers are seen near the top of the Co layer and the Co/Si interface. Adjacent to the high density layer near the Co/Si interface, a lower density Co with higher magnetic moment

consistent with theory (Fig. 4a) is seen. This measurement and that in Fig. 4 are on different parts cut from a wafer sample. (b) Spin asymmetry (SA) data [defined as: $(R^+ - R^-)/(R^+ + R^-)$, where $R^\pm$ are spin dependent reflectivity ] and the theoretical fit corresponding to the SLDs in the inset of (a). (c) SA data and fit to a uniform nuclear and a uniform magnetic SLD, as shown in (e). (d) SA data and fit to a uniform nuclear SLD and a nonuniform magnetic SLD, as shown in (f). In both (c) and (d) the fit is inconsistent with the data.

Because of low neutron intensity at the DHRUVA facility, the PNR data could be obtained only up to a small Q-range, we decided to make measurements also at the Magnetism Reflectometer [14] at the Spallation Neutron Source, ORNL facility, up to a much higher Q to obtain a higher spatial resolution. These results are shown in Fig. 5. The nuclear and the magnetic SLDs (in the inset of Fig. 5a), broadly have similar nature with those in Fig.4 suggesting higher NSLD with no ordered ferromagnetism in Co near the air/Co and the Co/Si interfaces. There are some differences in the details of the depth profiles of the nuclear and the magnetic SLDs compared to the results obtained from the measurement at DHRUVA. The reasons for the difference are twofold. Firstly, these results are not from the same part of the sample. The Co layer was grown on a Si wafer. The sample used for XRR experiment and PNR experiment at DHRUVA and the one for PNR experiment at ORNL are from different parts of the film, although cut from the same wafer. Another reason is, due to the higher Q range covered at SNS we have better spatial resolution for a refinement of the parameters for fitting the PNR data. The fit of ORNL data suggests small NSLD variation within the Co (interior or bulk region) layer with variation in magnetization. As we see from the plan-view TEM images (Fig. 3a,b), the density of the Co layer is not laterally uniform. Different grains of the polycrystalline Co film have different densities. XRR and PNR experiments have provided laterally averaged information from different parts of the film. So some discrepancy in the results is expected. From Fig. 5 we notice a region of the Co film near the Co/Si interface where the density of Co (nuclear SLD) is somewhat lower than that of normal Co. The corresponding magnetic SLD in this region is also marginally higher. This is consistent with the theoretical prediction, as seen in Fig. 4(a), Fig. 5 (b) show the spin asymmetry (SA) data defined as $(R^+ - R^-)/(R^+ + R^-)$, where $R^\pm$ are spin dependent reflectivity [8, 9, 10, 13]. For comparison we have also assumed different NSLD and MSLD profiles to fit the PNR data. Fig. 5 (c) and (d) are SA data and fits assuming two different cases, (i) uniform NSLD and MSLD depth profiles as shown in (e) and (ii) uniform NSLD and non-uniform MSLD profiles as shown in (f), respectively. It is evident from Fig. 5 (c-f) that these models do not fit the PNR (SA) data. Thus all the three reflectometry results, XRR, PNR at Dhruva and at ORNL clearly confirm the existence of a higher density Co layer at air/film and film/substrate interface with nearly zero magnetic moment density.

The enhanced density of Co has been observed near the top surface and at the Co/Si interface of the Co film. At the beginning of Co deposition on Si, due to higher surface free energy of Co compared to Si, Co would tend to grow as island. When islands coalesce and further deposited atoms enter into grain boundaries, they would generate a compressive stress on the Co grains thereby enhancing the material density [7]. Thereafter, the growth is Co on already deposited Co layer. The polycrystalline Co growth would continue. If the grain density increases near the top and more atoms enter into these grain boundaries, that would increase the density near the top of the film. However, to understand the mechanism of the formation of superdense phase of material in thin films, detailed investigations depending on growth conditions, such as deposition rate, growth temperature etc., are necessary. Similar enhancement effect, both at the interface and near the top of a thin film, has been observed in other systems, which could be understood in terms of surface free energy differences [15].

**Conclusions**

Depth profiles of electron scattering length density from XRR, as well as nuclear scattering length density from PNR, in a polycrystalline cobalt film on silicon have revealed regions of cobalt with much higher density compared to normal cobalt. These regions are near the top of the cobalt film and at the Co/Si interface. This finding is also corroborated by SIMS and TEM results. The density of the dense layer near the top of the film is about 1.3-1.4 times that of normal Co and its structure is fcc; below this layer the film is hcp Co of usual density. As revealed by the depth profile of the magnetic scattering length density, the superdense Co layer is nonmagnetic. Usually such a high density phase can only be formed under high pressure. However, during the growth of a polycrystalline film via atomic deposition, a compressive residual stress on polycrystalline grains due to incorporation of atoms in the grain boundaries can form high density material. Apparently, that is how this high density layer has formed. The evidence for nonmagnetic superdense Co indicates the possibility of existence of nonmagnetic Co in the earth's core.

**Acknowledgement**

We acknowledge the help of Jay Krishna, Dr. Avijit Das and Anjan Bhukta in XRR, SIMS and RBS experiments respectively. We thank the staff at the Ion Beam Laboratory, Institute of Physics, Bhubaneswar. Helpful discussion with Dr. Sumalay Roy is gratefully acknowledged. The work has been partially supported by the IBIQuS project. N.B. is supported by CSIR fellowship (09/080(0765)/2011-EMR-I). The work performed at SNS ORNL was supported by the Scientific User Facilities Division, Office of Basic Energy Sciences, DOE.

**Experimental methods**

A thin cobalt film (25 nm) was deposited on piranha cleaned, HF-etched Si(111) substrate in high vacuum by electron-beam evaporation method. Then the cobalt film is taken out of the vacuum chamber. The exposure of the film to air led to some surface oxidation. X-ray reflectometry (XRR) experiment was carried out with Cu K$_\alpha$ X-rays. We have carried out polarized neutron reflectivity (PNR) experiment on this cobalt film using the neutron reflectometer in DHRUVA, Bhabha Atomic Research Centre, Mumbai, India, which uses neutrons of wavelength 2.5Å [13]. Because of low neutron intensity at the DHRUVA facility, the accessible momentum transfer (Q) range is quite small. PNR experiment was repeated at the Oak Ridge National Laboratory facility where the accessible momentum transfer range is large. Rutherford backscattering spectrometry (RBS), secondary ion mass spectrometry (SIMS) and cross sectional transmission electron microscopy (XTEM) experiments were carried out in order to remove any ambiguity in the interpretation of the PNR and XRR data.

XRR and PNR are non-destructive techniques from which the depth dependent structure of the sample with nanometer resolution averaged over the lateral dimensions of the entire sample (typically 100 mm$^2$) can be inferred [8-10]. XRR and PNR involve measurement of the x-ray/neutron radiation reflected from a sample as a function of wave vector transfer *Q* (i.e., the difference between the outgoing and incoming wave vectors). In case of specular reflectivity (angle of incidence = angle of reflection) $Q = \frac{4\pi}{\lambda} sin\theta$, where θ is the angle of incidence and λ is the wavelength of x-ray/neutron, and it is qualitatively related to the square of the Fourier transform of the scattering length density (SLD) depth profile $\rho(z)$ (normal to the film surface or along the z-direction) [8-10]. For XRR, $\rho_x(z)$ is proportional to electron density whereas for PNR, $\rho(z)$ consists of nuclear and magnetic SLDs such that $\rho^\pm(z) = \rho_n(z) \pm CM(z)$, where *C* = 2.9109×10$^{-9}$ Å$^{-2}$ m/kA, and *M*(z) is the magnetization (kA/m) depth profile [8-10]. The sign +(-) is determined by the condition when the neutron beam polarization is parallel (opposite) to the applied field and corresponds to reflectivities $R^\pm$.

RBS experiment was carried out with 1 MeV He$^+$ ions. Scattered ions were detected at a scattering angle of 165 degree. RBS can detect the presence of any high-Z element with very high sensitivity (abouth a hundreth of an atomic layer). RBS is a non-destructive technique. SIMS experiments were carried out by sputtering the sample with 3 keV Ar+ ions at an ion beam current of 15 nA and detecting various species of sputtered molecular ions. Low beam current density (beam current/rastered area) was chosen to ensure low sputter-erosion rate so as to improve the SIMS depth resolution. As a function of time the sputtered ions come from different depths thereby providing the depth profile. TEM measurements were carried out with a 200 keV electron beam.